 \documentclass[prl,amsmath,amssymb,twocolumn, showpacs, superscriptaddress,10pt]{revtex4-1}

\usepackage{amsmath}
\usepackage{hyperref}
\usepackage{graphicx}
\usepackage{amsfonts}
\usepackage{amsthm}
\usepackage{cases}
\usepackage{bm}

\usepackage{color}
\definecolor{Blue}{rgb}{0.00, 0.00, 1.00}
\definecolor{Red}{rgb}{1.00, 0.00, 0.00}

\hypersetup{
    colorlinks=true,       
    linkcolor=red,          
    citecolor=blue,        
    filecolor=magenta,      
    urlcolor=cyan           
}

\newcommand{\nn}{\nonumber}
\newcommand{\be}{\begin{equation}}
\newcommand{\ee}{\end{equation}}
\newcommand{\bea}{\begin{eqnarray}}
\newcommand{\eea}{\end{eqnarray}}

\newcommand{\beq}{\begin{equation}}
\newcommand{\eeq}{\end{equation}}
\newcommand{\beqn}{\begin{eqnarray}}
\newcommand{\eeqn}{\end{eqnarray}}

\DeclareMathOperator{\Ai}{Ai}

\DeclareMathOperator{\Det}{Det}

\begin{document}

\title{{Multicritical} edge statistics for the momenta of fermions in non-harmonic traps}

\author{Pierre Le Doussal}
\affiliation{CNRS-Laboratoire de Physique Th\'eorique de l'Ecole Normale Sup\'erieure, 24 rue Lhomond, 75231 Paris Cedex, France}
\author{Satya N. \surname{Majumdar}}
\affiliation{LPTMS, CNRS, Univ. Paris-Sud, Universit\'e Paris-Saclay, 91405 Orsay, France}
\author{Gr\'egory \surname{Schehr}}
\affiliation{LPTMS, CNRS, Univ. Paris-Sud, Universit\'e Paris-Saclay, 91405 Orsay, France}

\date{\today}

\begin{abstract}
We compute the joint statistics of the momenta $p_i$ of $N$ non-interacting 
fermions in a trap, near the Fermi edge, with a particular focus on the largest one $p_{\max}$.
For a $1d$ harmonic trap, momenta and positions play a symmetric role and 
hence, the joint statistics of momenta is identical to that of the 
positions. In particular, $p_{\max}$, as $x_{\max}$, is distributed according to the Tracy-Widom distribution. 
Here we show that novel ``momentum edge statistics'' emerge when the curvature of the potential vanishes, 
i.e. for "flat traps" near their minimum, with $V(x) \sim x^{2n}$ and $n>1$. These are based on generalisations of the Airy kernel
that we obtain explicitly. The fluctuations of $p_{\max}$ are governed by
new universal distributions determined from the $n$-th member of the second Painlev\'e hierarchy of non-linear differential equations, {with connections to multicritical random matrix models.} Finite temperature extensions and possible experimental signatures in cold atoms are discussed.
\end{abstract}



\maketitle

{Fermi gases in confining traps exhibit an edge in space,
where the density vanishes \cite{GPS08,BDZ08,castin}. As a result, the 
quantum and thermal fluctuations are greatly enhanced
leading to interesting {{edge}} physics.} The case of spinless non-interacting fermions
is more tractable analytically: recently a number of
theoretical predictions, at zero and finite temperature, 
were obtained \cite{Kohn,Eis2013,us_finiteT,DPMS:2015,fermions_review,Dubail}. These are ripe to be tested in cold atom experiments 
where the non-interacting limit can be reached \cite{Chin2010,2DBoxGaz}. Two complementary 
experiments are possible, either the quantum microscopes, 
which measure jointly the {\it positions} $x_i$ of the fermions 
\cite{Cheuk:2015,Haller:2015,Parsons:2015,Omran2015},
or the time of flight experiments which measure jointly {their}
{\it momenta} $p_i$ \cite{TOF,2DBoxGaz}. 

Most of the recent predictions concern the {\it positions} $x_i$ of $N$ noninteracting
fermions. 
At zero temperature, $T=0$, in one dimension, $d=1$, and for a harmonic trap, $V(x) = \frac{1}{2} m \omega^2 x^2$,
they are in one-to-one correspondence \cite{Eis2013,us_finiteT,marino_prl,CMV2011,CDM14,castillo} with the eigenvalues of a complex Hermitian matrix
with independent Gaussian entries, known as the Gaussian Unitary Ensemble (GUE) of random matrix theory \cite{mehta,forrester}. Their joint probability distribution
function (PDF) is given by the Wigner-Dyson formula. 
The fermion density is thus the Wigner semi-circle
which vanishes at the edge $x_e$ as $\rho(x) \sim |x-x_e|^{1/2}$. Consequently, the spatial quantum correlations at the edge of the trap are described by the fluctuations of the few largest eigenvalues of the GUE~\cite{fermions_review}. These form a {\it determinantal point process}
\cite{johansson,borodin_determinantal,tracy_widom_determinantal}, i.e all correlation functions can be written as determinants with entries called kernel.
\begin{figure}[hb]
\includegraphics[width = \linewidth]{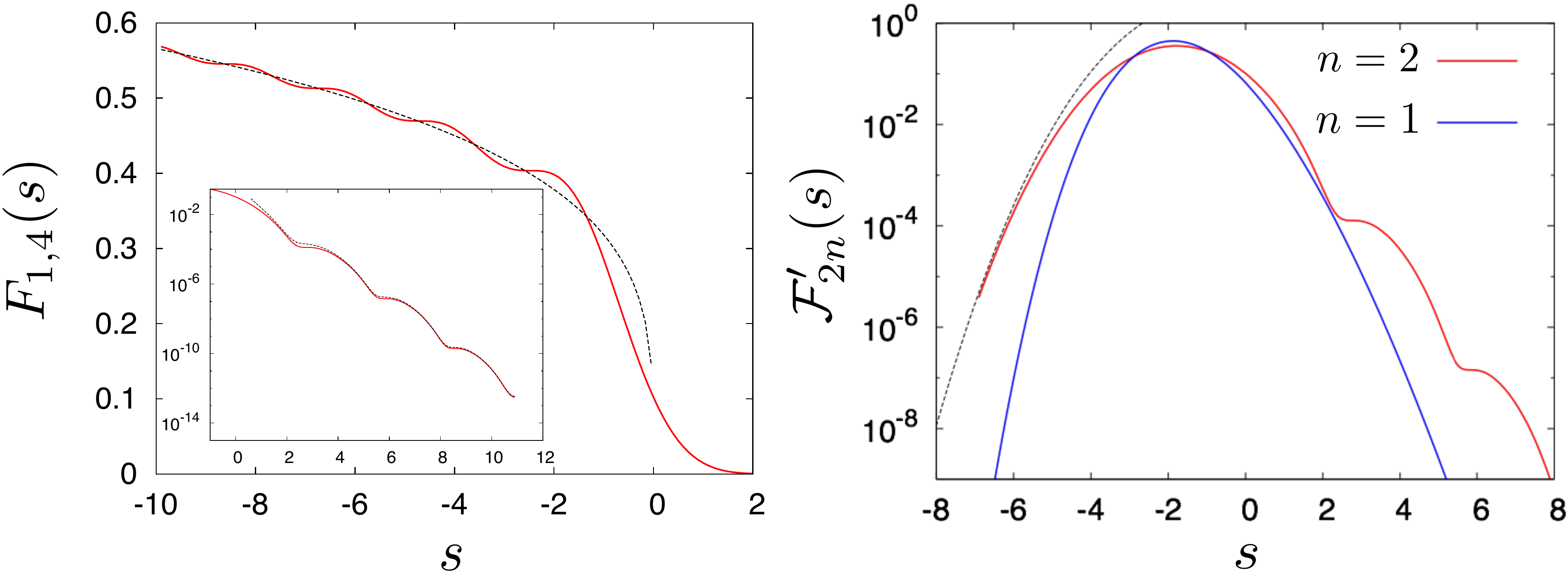}
{\caption{{\it Left:} Solid line: scaling function $F_{1,2n=4}(s)$, Eq. \eqref{densityedge},
of the density at the edge, for $V(x) \sim x^4$. Dotted line: $s \to -\infty$ asymptotics
$F_{1,4}(s) \simeq \frac{|s|^{1/4}}{\pi}$, matching the bulk density, see 
above~\eqref{densityedge}. Inset: same in log-linear scale showing the large $s>0$ oscillations. 
Dotted line: asymptotics in~(\ref{righttail}). {\it Right}: Scaling function ${\cal F}'_{2n}(s)$ of the PDF of $p_{\max}$ in (\ref{scalF})-(\ref{F2nint}). Red: $n=2$. Blue: the TW distribution ($n=1$) for comparison.
Dotted line: large $s<0$ behavior, ${\cal F}'_{4}(s) \sim \exp(- \frac{\sqrt{8/3}}{15} |s|^{5/2})$.}\label{Fig_density}}  
\end{figure}
In the case of GUE, this kernel is called the Airy kernel \cite{TracyWidomAiry}, as
it is related to the Green's function of a single quantum particle in a linear
potential.
In particular the PDF of the (properly centered and scaled) position of the rightmost fermions $x_{\max}= \max_{i=1,\ldots,N} x_i$
is given \cite{us_finiteT} 
by the celebrated GUE Tracy Widom (TW) distribution \cite{TracyWidomAiry}, which also arises in 
many 
problems in mathematics and physics \cite{baik,johann,growth,CLR10,DOT10,ACQ11,sequence,MS_thirdorder} and was
measured in experiments \cite{takeuchi,davidson,lemarie} in other contexts. 
These properties were shown to extend to finite temperature, in terms of a one-parameter deformation of the Airy kernel, indexed
by the reduced temperature $b= \hbar \omega N^{1/3}/T$ (with a corresponding finite $T$ extension of the TW distribution) \cite{us_finiteT,fermions_review,LiechtyT}.
Remarkably, these spatial edge correlations were shown to be universal, 
independent of the details of the smooth confining potential \cite{Eis2013,fermions_review}, which can be traced to
the fact that the density at the edge still vanishes as $|x-x_e|^{1/2}$.
Extensions to dimensions
$d>1$ ~\cite{DPMS:2015,farthest_f,torquato} {{and non-smooth potentials, e. g. hard box \cite{FFGW03,Cunden1D,UsHardBox},
 were also studied}}.

One can now ask about the statistics of the {\it momenta} $p_i$ of $N$ non interacting
fermions, and their maximum $p_{\max}= \max_{i=1, \ldots,N} p_i$, in a (e.g. $1d$) trap described by a single particle Hamiltonian $H=\frac{p^2}{2 m} + V(x)$. 
It is an important question to make predictions for 
time of flight experiments in traps 
of varying shapes, as can be currently designed \cite{BDZ08,Zwi2017,2DBoxGaz}.
If the potential is bounded from below, there exists also an edge in momentum space
$p_e$, beyond which the momentum density vanishes. Obviously, if the confining potential is harmonic, momenta and positions play a
symmetric role and the two (dimensionless) random sets
$\{p_i/\hbar \alpha \}_{i=1,\ldots,N}$ (momenta) and $\{\alpha x_i\}_{i=1,\ldots,N}$ (coordinates) 
are described by exactly the same joint PDF (here $\alpha = \sqrt{m \omega/\hbar}$
is the harmonic oscillator inverse length scale), at any temperature (and, in fact, in any $d$). 
The question of what happens for a more
general, non harmonic trap, $V(x) \sim x^{2n}$ is however, non-trivial and open. 
Interestingly, as we find below, the density {\it in momentum space} vanishes as
\be \label{density} 
\rho(p) \sim (p_e-p)^{\frac{1}{2 n}} \;,
\ee 
i.e, distinct from the standard Wigner 
semi-circle exponent $\frac{1}{2}$ (for $n=1$)
which suggests a new universality class.
In particular we expect that the fluctuations of $p_{\max}$ 
are given by a distribution different from the TW.

On the random matrix side, there exist generalizations of the GUE
involving matrix potentials such that the density of eigenvalues
vanishes {{at the edge}} with a rational exponent ${\sf q}/{\sf p}$ \cite{ReviewMatrixModels}. In the so-called double
scaling limit \cite{gross1990nonperturbative,douglas1990strings,brezin1990exactly,BowickBrezin}, these matrix models exhibit multicritical points
indexed by two integers $({\sf p},{\sf q})$ with universal properties
\cite{BrezinHouches,KazakovDaul,Eynard,BrezinHikamiSource,janik,AkemanNew}.
Such models were introduced in 
the context of random surfaces and string theory \cite{BIPZ,GrossWitten,Periwal1,Douglas,Mandal1,Mandal2,Douglas2,Klebanov,marino}.
A special class of these models were recently studied
in \cite{Claes1,Gernot_TW}, leading to generalizations of the TW distribution. 
A natural question is whether there are experimentally relevant settings
where the universal physics near these multicritical points can be accessed.
Given the behavior of the density \eqref{density}, the momentum
statistics of 
fermions near the edge may be a natural candidate.

Indeed, in this paper we demonstrate that the momentum statistics at the edge of noninteracting
fermions in some non-harmonic potentials (i) leads to universality classes
different from the GUE-TW class
(ii) appears to be in 
correspondence with 
a class of
multicritical matrix models. 
We focus on $d=1$ and
consider potentials which (with no loss
of generality) attain their minimum 
at $x=0$, i.e. $V(0)=V_{\min}=\min_x V(x)=0$.
Two edges in momentum space then exist at 
$\pm p_e$ with $p_e =\sqrt{2 m \mu}$, $\mu$ being the Fermi energy. We show that the set of momenta $\{ p_i \}_{i=1, \ldots,N}$ forms a determinantal point process, which, near
the edges, is characterized by universal kernels. The universality classes however depend on the
behavior of the potential near its minimum. If the curvature there is non-zero, the
universality class is the one of the Airy kernel (e.g. leading to the TW distribution for properly centered and scaled $p_{\max}$).  
If the curvature vanishes, e.g. if $V(x) \sim x^{2n}$ with $n>1$, which is the case
e.g. for the simplest pure power law confining potentials 
\be
H = \frac{p^2}{2 m} + g \, x^{2 n}  \quad , \quad g>0 \label{pure}
\ee
then we show that there exists one new universality class for each integer $n>1$.

Let us now describe our main results, specializing to
the pure $x^{2 n}$ potentials \eqref{pure}, generalizations are discussed later.
We start with $T=0$. First, the momentum density $\rho(p)=
\langle \sum_{i=1}^N \delta(p-p_i) \rangle$ in the bulk is found to be \cite{SM}
$\rho(p) \simeq \frac{1}{\hbar \pi} [\frac{1}{g} (\mu - \frac{p^2}{2 m})]^{\frac{1}{2 n}}$
for $|p|<p_e=\sqrt{2 m \mu}$ 
for large $N=\int dp \rho(p)$, which as discussed above vanishes 
as in~\eqref{density}. For large but finite $\mu$ and $N$ the density
is smeared out near the edge $p \approx p_e$
over a layer of width~\cite{footnote0}
\be
p_N = \hbar \left(\frac{m g}{\hbar p_e}\right)^{\frac{1}{2 n+1}} \sim \mu^{- \frac{1}{4n+2}}
\label{pN}
\ee and
takes the scaling form
\be \label{densityedge} 
\rho(p)  \simeq  \frac{1}{p_N} F_{1,2n}(\frac{p-p_e}{p_N}) \, , \, F_{1,2n}(s) = \int_s^{\infty} du \Ai_{2 n+1}(u)^2,
\ee 
{{where ${\rm Ai}_{2n+1}(x)$ is defined by its integral representation
\be
\Ai_{2 n +1}(x) = \int_{i \mathbb{R} + \epsilon} \frac{dz}{2 i \pi} e^{ (-1)^{n} x z +  \frac{1}{(2 n+1)} z^{2 n+1} }
\label{Ain} 
\ee 
where $\epsilon= (-1)^{n+1} 0^+$. This is the real solution of $\partial_x^{2n} {\rm Ai}_{2n+1}(x)= 
(-1)^{n+1} x {\rm Ai}_{2n+1}(x)$. For $n=1$ (harmonic potential), ${\rm Ai}_3(x)\equiv {\rm Ai}(x)$ is the 
standard Airy function. Eq.~(\ref{Ain}) then provides a generalisation for~$n>1$.}}
%
In Fig. \eqref{Fig_density}
we show a plot of $F_{1,4}$ whose asymptotics read~\cite{SM}
$F_{1,4}(s) \simeq \frac{1}{\pi} |s|^{1/4}$ for $s \to -\infty$,
exactly matching the density in the bulk, and 
\be \label{righttail} 
F_{1,4}(s) \simeq \frac{1}{8 \pi s} e^{- \frac{4}{5} \sqrt{2} s^{5/4}} \left(\sqrt{2} + \cos( 
\frac{4}{5} \sqrt{2} s^{5/4}) \right) \,  
\ee
for $s \to + \infty$,
which exhibits, contrary to the standard harmonic oscillator $n=1$, 
non trivial oscillations (see Fig. \ref{Fig_density}). These can be understood
from \eqref{pN}, which shows that the fluctuation scale $p_N$ 
{\it increases} with increasing $n$ at large $\mu$.
In addition we show that the rescaled fermion momenta near the edge
$k_i = \frac{p_i-p_e}{p_N}$
form a determinantal point process (see definition below)
characterized by the~kernel
\be
K_{\Ai,2 n+1}(s,s')= \int_0^{+\infty} du \Ai_{2n+1}(s+u) \Ai_{2n+1}(s'+u) \label{Kn} 
\ee
which generalizes the Airy kernel (for $n=1$). 
This implies that the cumulative distribution function (CDF) of the largest momentum $p_{\rm max}$ 
takes the scaling form
\be
{\rm Prob}( p_{\rm max} < p) \simeq {\cal F}_{2n}\left(\frac{p-p_e}{p_N}\right) \label{scalF} 
\ee
where the universal scaling function is given by the Fredholm determinant (FD)
\cite{fredholm}
\be
{\cal F}_{2n}(s) = \Det [ I - P_s K_{\Ai,2n+1} P_s ] \label{calF} 
\ee
{{where $P_s$ is a projector onto the 
interval $[s,\infty[$. For $n=1$ (harmonic potential) it  
reduces to the celebrated GUE-TW distribution ${\cal F}_{2}(s) =F_2(s)$.}}
In that case it is well known that this FD 
can be obtained from the solution of the
Painlev\'e II equation
\cite{TracyWidomAiry}. Here we obtain a
more general result, i.e. that ${\cal F}_{2n}(s)$ for any $n$ can be expressed~as
\be
\ln {\cal F}_{2 n} (s) = - \int_{s}^{+\infty} du (u-s) q(u)^2 \label{F2nint}
\ee 
where $q(s)$ satisfies a non-linear differential equation known as
the $n$-th member of the second Painlev\'e hierarchy (see e.g. \cite{Claes1}), denoted as $P_{\rm II}^{(n)}$, 
for some specific values of the parameters, with boundary
condition at $s \to +\infty$ given by $q(s) \sim_{s \to +\infty} \Ai_{2 n+1}(s)$
\cite{SM}.
For $n=1$ it is the standard Painlev\'e II equation $q''=s q + 2 q^3$
and Eq. \eqref{F2nint} leads to the well-known TW distribution.
We give here the second (fourth order) nonlinear equation which holds for $n=2$
\bea
q'''' = 
q \left(10 (q')^2+10 q q''-6
   q^4-s\right) \label{PII2new} 
\eea
It allows to plot (see Fig. \ref{Fig_density}) the PDF,
${\cal F}'_{4}(s)$, for the centered and scaled $p_{\max}$ \cite{footnotesigma}. {{It
also allows to extract the asymptotics 
${\cal F}_{2n}(s) \sim \exp( - f_n (-s)^{2 + \frac{1}{n}}  )$ 
for $s \to - \infty$,}}
with $f_n =
\frac{n^2}{4(2 n+1)(n+1)} [ \frac{\sqrt{\pi} \Gamma(n+1)}{\Gamma(n + \frac{1}{2})} ]^{\frac{1}{n}}$.
For large $s>0$, ${\cal F}_{2n}'(s) \simeq F_{1,2n}(s)$ which, for $n=2$, is
given in \eqref{righttail}, exhibiting a striking oscillatory behavior. Interestingly, the same Painlev\'e hierarchy also appears in 
multicritical matrix models, as discussed below. 

We then extend these results to finite temperature. We find that
the temperature window to observe these anomalous edge behavior
is $T \sim  \mu^{\frac{n}{2 n+1}}$. In this regime we obtain a
modified kernel \eqref{kernelT} and scaling function for the
density near the edge, 
depending only on $n$ and on the 
scaled  inverse temperature parameter $b$ 
defined in
\eqref{defb}. Finally we discuss universality of our results with respect
to the form of the kinetic energy and the potential.

To establish these results, we first study the single particle eigenstates. We denote $\epsilon_k$
the eigenenergies (in increasing order) of $H$ in \eqref{pure}, with $k \geq 1$ an integer label,
and $\psi_k(x)$ the corresponding eigenfunctions in real space. The eigenfunctions in
momentum space, $\hat \psi_k(p)= \int
dx e^{- i p x/\hbar} \psi_k(x)$, 
obey the eigenvalue equation
\be
\left( \frac{p^2}{2 m} + g   (\hbar i \partial_p)^{2 n} \right) \hat \psi_k(p) = \epsilon_k \hat \psi_k(p) \label{eqp} 
\ee
using {{the representation $x \equiv i \hbar \partial_p$}} in the momentum basis. 
We now study how these momentum space wavefunctions behave near the edge
at $p=p_e=\sqrt{2 m \mu}$. We write $p = p_e + \tilde p$ and the linearized version of \eqref{eqp}
reads
\bea
\left( \frac{p_e \tilde p}{m} + g (\hbar i \partial_{\tilde p})^{2 n} \right) \hat \phi_k(\tilde p) = (\epsilon_k-\mu)  \hat \phi_k(\tilde p) \label{eqp2} 
\eea 
where $\hat \psi_k(p) = \hat \phi_k(\tilde p)
= \int dx e^{ - i \tilde p x/\hbar} \phi_k(x)$ and $\phi_k(x) = e^{- i p_e x/\hbar} \psi_k(x)$. We are discarding terms of $O(\tilde p^2)$, which amounts to neglect $\tilde p \sim p_N \ll p_e$, where $p_N$ is given in \eqref{pN}. Eq. \eqref{eqp2} can be solved by going back to real space
\bea
\left( - i \frac{\hbar p_e}{m} \partial_x + g \, x^{2 n} \right) \phi_k(x) = (\epsilon_k-\mu) \phi_k(x) \label{eqp2} 
\eea 
whose solution is
\bea
\phi_k(x)  \sim e^{i  \frac{m(\epsilon_k-\mu)}{\hbar p_e} x - i \frac{m g}{(2 n+1) \hbar p_e} x^{2 n+1} }
\eea 
Going back to momentum space, setting $x = (-1)^n i z (\frac{\hbar p_e}{m g})^{\frac{1}{2 n+1}}$
and comparing with \eqref{Ain}, we obtain that the eigenfunctions take the 
form
near the edge 
\be
 \hat \psi_k(p) \sim  \Ai_{2 n +1} \left(\frac{p-p_e}{p_N} +\frac{m(\mu-\epsilon_k)}{p_e p_N} \right)\;. \label{PsiAin} 
\ee

We now consider $N$ noninteracting fermions with single particle Hamiltonian
\eqref{pure}. The ground state wavefunction in momentum space, $\Psi_0(p_1,\ldots, p_N)$,
 is a Slater determinant constructed from
the $N$ eigenfunctions $\hat \psi_k$, $k=1, \ldots, N$ with lowest energies. 
As a result, the quantum probability can be expressed as a determinant
\be
|\Psi_0(p_1,\ldots, p_N)|^2 = \frac{1}{N!} \det_{1 \leq i,j \leq N}[ K_\mu(p_i,p_j) ] \label{Q}
\ee 
involving the kernel $K_\mu(p,p') = \sum_{k=1}^N \hat \psi_k^*(p) \hat \psi_k(p')$,
which is self reproducing \cite{footnoteSelf}.
This property implies \cite{mehta} that the set of $\{p_i\}_{i=1, \ldots, N}$, distributed
with the quantum probability \eqref{Q}, forms a determinantal 
point process. It implies that the $k$-point correlation
functions at $T=0$ 
\be
R_k(p_1, \ldots,p_k) = \frac{N!}{(N-k)!} \int \prod_{j=k+1}^N dp_{j} |\Psi_0(p_1,\ldots, p_N)|^2
\ee
can be written as determinants
\be
R_k(p_1,\ldots,p_k) = \det_{1 \leq i,j \leq k}[ K_\mu(p_i,p_j) ] \;.
\ee
In particular the density is $\rho(p):=R_1(p)=K_\mu(p,p)$. The 
full counting statistics for $N_J$, the number of fermions
in any subset $J$, is obtained from its Laplace transform as
$\langle e^{- r N_J } \rangle = \Det[I - (1-e^{-r}) P_J K_\mu]$
where $P_J$ projects on $J$ \cite{johansson,borodin_determinantal}. 
Choosing $J=[p,+\infty[$ in the limit
$r \to +\infty$, yields the 
CDF of the maximum momentum as (with $P_p:=P_{[p,+\infty[}$)
\bea
&& {\rm Prob}( p_{\rm max} < p) = \Det[I - P_p K_\mu P_p] \label{Fd} 
\eea
 
To study the correlations, or counting statistics, 
near the edge (e.g. the statistics of the few highest values of $p_i$) 
we use the edge behavior of the eigenfunctions as discussed above in \eqref{PsiAin}.
We now insert \eqref{PsiAin} into the kernel $K_\mu$, 
and use the continuous basis orthonormality
$\int_{-\infty}^{+\infty} dx {\rm Ai}_{2n+1}(x+s) {\rm Ai}_{2n+1}(x+s') = \delta(s-s')$.
We can then replace the discrete sum over $k$ in $K_\mu$ 
by an integral, 
using similar arguments as in \cite{Eis2013}. We
obtain that the kernel $K_\mu$ takes the
scaling form near the edge
\bea \label{Kscal} 
&& K_\mu(p,p') \simeq  \frac{1}{p_N} K_{\Ai,2 n+1}\left( \frac{p-p_e}{p_N} , \frac{p'-p_e}{p_N}\right) 
\eea 
where $K_{\Ai,2 n+1}$ is defined in \eqref{Kn}. For $n=2$ it reads 
\bea
  K_{{\rm Ai}, 5}(s,s') &=& \frac{1}{s-s'} [   \Ai'_{5}(s) \Ai''_{5}(s') - \Ai''_{5}(s) \Ai'_{5}(s') \nn
\\
&& - \Ai_{5}(s) \Ai'''_{5}(s') + \Ai'''_{5}(s) \Ai_{5}(s') ] \label{CD5} 
\eea
with similar expressions (with $2n$ terms) for any $n$, generalizing
the standard Airy kernel \cite{Pearcey}.
Explicit expressions in terms of hypergeometric
functions and asymptotic expansions are given in \cite{SM}.
Since $\rho(p)=K_\mu(p,p)$, this establishes
the result in \eqref{densityedge} and its asymptotics. From \eqref{Fd}, we find that the 
CDF of the maximum momentum, $p_{\max}$,
also takes the scaling form \eqref{scalF} 
with the scaling function ${\cal F}_{2n}(s)$ given by the Fredholm determinant \eqref{calF}.
We extended the method of calculation of \cite{TracyWidomAiry,BrezinHikamiSource}
to show that these FD can be written as in \eqref{F2nint} where $q$, together with 
a set of auxiliary functions, satisfy a system of non linear coupled first order differential equations.
Remarkably, $q$ can be shown to satisfy a closed differential equation, of order $2n$,
which, furthermore
identifies with the $n$-th member of the Painlev\'e $P^{(n)}_{\rm II}$ hierarchy, as given e.g. in \cite{Claes1}, see \cite{SM}. The case $n=1,2$ were discussed above.

Interestingly, the same Painlev\'e hierarchy $P^{(n)}_{\rm II}$ occurs in multicritical unitary random matrix models
\cite{Periwal1} defined by the partition function $Z = \int dU e^{ N{\rm Tr} \, V(U+U^+)}$
where the integral is over the unitary group $U(N)$. Here $V(z)$ is a polynomial
which by fine tuning leads to a sequence of multicritical points.
For $V(z) = g z$, there is a phase transition at infinite $N$
for $g=g_c$, between strong and weak coupling phases \cite{GrossWitten}.
In the double scaling limit $g-g_c \sim N^{-2/3}$ one finds that
the partition sum $Z$ is proportional to the GUE-TW distribution 
$F_2(s)$ with $s \sim N^{2/3}(g-g_c)$ and thus relates to 
the standard Painlev\'e II equation. 
Interestingly, for appropriate polynomials $V(z)$ of degree $n$, one finds \cite{Periwal1}
multicritical points in the double scaling limit
$g-g_c \sim N^{-2n/(2n+1)}$, where $Z$ is now related to the $n$-th member
of the $P_{\rm II}^{(n)}$ hierarchy. Similar multicritical behavior arise for Hermitian matrices \cite{Eynard,Douglas2,Douglas} and belong to the
universality class $({\sf p},{\sf q})$
(e.g. $(2,4 n)$ in \cite{Periwal1,Klebanov})
with a density vanishing at the edge with exponent $2 n$. Remarkably,
there is a {\it duality} between $({\sf p},{\sf q}) \Leftrightarrow ({\sf q},{\sf p})$
worked out in two-matrix models 
\cite{KazakovDaul,ReviewMatrixModels} (the two models share the same
partition function). It is tempting to conjecture that the universality class found here is
related to
one of the $({\sf p},{\sf q})$
multicritical matrix models, with a
density 
exponent $1/(2 n)$
corresponding to a strong to weak coupling transition of order $2+ \frac{1}{n}$, 
generalizing the third order transition for $n=1$ \cite{MS_thirdorder}.
Interestingly, 
simple realizations of multicritical
Gaussian matrix models in presence of a source
lead to a density that vanishes at the edge with 
an exponent $1/k$. 
The case $k=3$ 
\cite{BrezinHikamiSource} 
yields
the so-called Pearcey kernel \cite{Pearcey}.
We surmise that this class is related to the model of noninteracting
fermions studied here, for potentials 
$V(x) \sim |x|^{k}$ for any integer $k$ (even or odd). 

We now extend our study to finite temperature. In current experiments
one can prepare $1d$ tubes of cold noninteracting fermions
with $T/\mu \approx 0.3$ where $\mu$ is the $T=0$ Fermi energy. 
One defines the (dimensionless) 
reduced inverse temperature
\be
b = T^*/T \quad , \quad 
T^* = g^{\frac{1}{2 n+1}}  (2\mu \hbar^2/m)^{\frac{n}{2 n+1}} \label{defb} 
\ee 
and consider the temperature regime $T \sim \mu^{\frac{n}{2 n+1}}$ such that $b=O(1)$. 
Using the equivalence, for local observables, of the canonical and grand canonical
ensembles, we work in the latter, where the set of scaled and centered fermion momenta
$k_i=\frac{p_i-p_e}{p_N}$ form a determinantal process with associated kernel \cite{SM}
\be \label{kernelT} 
K_{2n+1,b}(s,s')= \int_{-\infty}^{+\infty} du \frac{\Ai_{2n+1}(s+u) \Ai_{2n+1}(s'+u) }{1 + e^{-b u}}.
\ee 
This leads to the density $\rho(p) = \frac{1}{p_N} F_{1,2n,b}(\frac{p-p_e}{p_N})$ near the edge
with the scaling function $F_{1,2n,b}(s)=K_{2n+1,b}(s,s)$.
%
It decays exponentially at large $s$, $F_{1,2n,b}(s) \simeq C_b e^{-b s}$, where
$C_b=\int du \Ai_{2n+1}(u)^2 e^{b u}$. Both the kernel \eqref{kernelT} and the density 
depend continuously on the scaled inverse temperature $b$, and
reduce to the $T=0$ results~\eqref{densityedge} and \eqref{Kn}
in the limit $b \to +\infty$. The PDF of the scaled and centered maximum momentum 
at finite temperature, $k_{\rm max}(T) =
(p_{\rm max}(T)-p_e)/p_N$ is given by the Fredholm determinant \eqref{calF} 
replacing $K_{\Ai,2n+1}$ by $K_{2n+1,b}$. {As $T$ is lowered,
the PDF of $k_{\rm max}(T)$ exhibits a universal crossover from a Gumbel to 
${\cal F}_{2n}$ distribution (of variance $\sigma_{2n}^2$). We propose to
check in experiments the variance
${\rm Var}[k_{\rm max}(T)] = (\frac{T}{T^*})^2 {\cal V}_{2n}(\frac{T^*}{T})$
with {{${\cal V}_{2n}(z)$}} a universal function, 
${\cal V}_{2n}(0)=\frac{\pi^2}{6}$ and ${\cal V}_{2n}(z) \sim \sigma_{2n}^2 z^2$ for large $z$.}


Our main results, i.e. the scaling forms \eqref{densityedge}-\eqref{scalF}-\eqref{kernelT},
are universal, with the same scaling functions, for (i) a larger class of potentials such that $V(x) \simeq g x^{2 n}$ near its (single \cite{footnotePeriodic}) minimum (ii) a more general kinetic energy $E(p)$, with
$p_N = \hbar (\frac{g}{\hbar E'(p_e)})^{\frac{1}{2 n+1}}$. Hence only the
behavior of $V(x)$ near its minimum determines the universality class,
which does not even require a confining trap and an edge in real space.
While for $n=1$ there is universality in the $(x,p)$ plane, as
can be seen from the study of the Wigner function \cite{UsWigner}, for $n>2$ the momentum and real space
edge physics are unrelated.

It is also natural to ask about the imaginary time quantum dynamics of the Hamiltonian (\ref{pure}).
The multi-time correlations of the centered and scaled
momenta can be expressed as determinants involving an {\it extended kernel}, 
here given by Eq. (99) in \cite{UsPeriodic} upon replacing the Airy functions by the $\Ai_{2n+1}$
functions in \eqref{Ain}. Thereby, in the $T=0$ limit, we obtain a generalization (to $n>1$) of
the celebrated Airy$_2$ process ($n=1$) \cite{PraSpo02,QR14}, 
as the imaginary time trajectory of the maximum 
momentum.



In conclusion we have unveiled new universality classes for
edge statistics of the momenta of fermions in $d=1$ non-harmonic traps,
which we hope can be measured in cold atom time of flight experiments.
We found unexpected connections to multicritical matrix models.
Given the ubiquity of the TW distribution, it would be of great interest to
find other physical systems where its generalized (multicritical) version, ${\cal F}_{2n}$, appears.
Another direction to be investigated
is the case of higher dimensional anharmonic potentials, 
such as $V(x,y)=(x^2+y^2)^2$
or $V(x,y)=x^2 + y^4$ in $d=2$, for which we expect \cite{footnoteExpect}
anomalous momentum edge behavior.
 


%
%
%
%
%
%
%



{\it Acknowledgments:} We thank D. S. Dean for useful discussions and ongoing collaborations.
We also thank D.~Bernard, E.~Br\'ezin, T. ~Claeys, B.~Eynard, V.~Kazakov and
and A. Krajenbrink for enlightening discussions. 
This research was partially supported by ANR grant ANR-17-CE30-0027-01 RaMaTraF.

{}

\newpage


\newpage

\begin{widetext} 

\bigskip

\bigskip

\begin{large}
\begin{center}

Supplementary Material for {\it 
Multicritical edge statistics for the momenta of fermions
in non-harmonic traps}

\end{center}
\end{large}

\bigskip

We give the principal details of the calculations described in the main text of the Letter. 

\bigskip

%
%
%
%
%
%
%
%


\medskip
\begin{center}
{\bf A. Density in momentum space}
\end{center}
\medskip

To compute the density in momentum space it is useful to start from the so-called
Wigner function $W_N(x,p)$, which is the analogous of a joint density in phase space
$(x,p)$. In particular by integrating over $x$ one obtains the momentum density (normalised to $N$) 
as $\rho(p) = \int dx W_N(x,p)$. In the large $N$ limit, and for an arbitrary potential $V(x)$ in $d=1$,
the Wigner function reads \cite{castin,UsWigner} (see definitions there)
\be
W_N(x,p) \simeq \frac{1}{2 \pi \hbar} \theta\left(\mu - \frac{p^2}{2 m} - V(x)\right)
\ee 
where $\theta(x)$ is the Heaviside step function. For the pure power law potentials,
$V(x)=g x^{2n}$ one finds
\be \label{dens1} 
\rho(p) \simeq \frac{1}{\hbar \pi} \left[\frac{p_e}{m g} (p_e-p)\right]^{\frac{1}{2 n}} 
\ee
where $p_e=\sqrt{2 m \mu}$. 
In that case one can relate $N$ to $\mu$ using $\int_{-p_e}^{p_e} dp \, \rho(p)=N$ which leads to
\bea
N \simeq \gamma_n \frac{p_e}{\hbar \pi} \left(\frac{p_e^2}{2 m g}\right)^{\frac{1}{2 n}}  = 
\gamma_n \frac{\sqrt{2 m}}{\hbar \pi} g^{- \frac{1}{2 n}} \mu^{\frac{n+1}{2 n}} = \frac{\gamma_n}{\pi} 
\left(\frac{\mu}{e_0}\right)^{\frac{n+1}{2 n}}  \label{Nmu} 
\eea 
which holds in the large $\mu$ limit. Here $\gamma_n := \int_{-1}^{+1} dz (1-z^2)^{\frac{1}{2 n}} = \frac{\sqrt{\pi} \Gamma(1+ \frac{1}{2 n})}{\Gamma(\frac{3}{2} + \frac{1}{2 n})}$. We have introduced the energy scale
$e_0=\frac{\hbar^2}{2 m} \alpha^2$ with $\alpha=(\frac{2 m g}{\hbar^2})^{1/(2 n+2)}$
the typical inverse length. For the harmonic oscillator $n=1$ one recovers $\mu \simeq N \hbar \omega$
and $e_0=\hbar \omega/2$. 

For more general potentials, but which behave as $V(x) \simeq g x^{2n}$ near its minimum,
the formula \eqref{dens1} is still valid, but only near the edge $p \approx p_e=\sqrt{2 m \mu}$,
leading to the same exponent $1/(2 n)$ as in \eqref{density}  for the behaviour of the density.
For more general potentials (e.g. with several scales) the relation \eqref{Nmu} between
$N$ and $\mu$ may be modified. The control parameter in this work is large $\mu$
(which implies also large $N$). 


\medskip
\begin{center}
{\bf B. Some properties of the $\Ai_{2n+1}$ functions}
\end{center}
\medskip
 
Let us establish some of the properties of the functions $\Ai_{2n+1}$. We recall the definition \eqref{Ain} 
in the text 
\be
\Ai_{2 n +1}(x) = \int_{C} \frac{dz}{2 i \pi} e^{ (-1)^{n} x z +  \frac{1}{(2 n+1)} z^{2 n+1} }
\label{Ain2} 
\ee 
where for the contour $C$ we can use $i \mathbb{R}$ with the understanding 
that it can be slightly deformed for absolute convergence. \\

{\bf Differential equation}.
Taking the $2 n$-th derivative of \eqref{Ain} we have
\bea 
&& \partial_x^{2 n} \Ai_{2 n+1}(x) = \int_{- i \infty}^{+ i \infty} \frac{dz}{2 i \pi} 
 z^{2 n} e^{ (-1)^{n} x z +  \frac{1}{(2 n +1)} z^{2 n +1} }
 = \int_{- i \infty}^{+ i \infty} \frac{dz}{2 i \pi} e^{ (-1)^{n} x z}
[ \partial_z e^{ \frac{1}{(2 n+1)} z^{2 n+1} } ]
 \\
&& = - \int_{- i \infty}^{+ i \infty} \frac{dz}{2 i \pi}  [ \partial_z e^{ (-1)^{n} x z} ]
 e^{ \frac{1}{(2 n +1)} z^{2 n +1} }
 = (-1)^{n+1} x \Ai_{2 n+1}(x) \nn
\eea 
using $(-1)^{2 n^2} =1$ in the first line. Hence we obtain the result given in text above Eq. \eqref{Ain} 
\be
\partial_x^{2 n} \Ai_{2n +1}(x) = (-1)^{n+1} x \Ai_{2 n+1}(x) \label{Aidiff} 
\ee
consistent with $\Ai''(x)=x \Ai(x)$ for $n=1$. \\

{\bf Orthonormality}. Next, let us show the continuous basis orthonormality and 
calculate the scalar product using the definition in \eqref{Ain} as
\bea
&& \int dx \Ai_{2 n +1}(x+s) \Ai_{2 n +1}(x+s') = \int dx 
\int_{-  \infty}^{+  \infty} \frac{dy_1}{2 \pi} 
\int_{-  \infty}^{+  \infty} \frac{dy_2}{2  \pi} 
e^{i (-1)^{n} ( (x+s) y_1 +  \frac{1}{(2 n+1)} y_1^{2n +1}  + (x+s') y_2 +  \frac{1}{(2n +1)} y_2^{2n +1} )} \nn \\
&& 
= \int_{-  \infty}^{+  \infty} \frac{dy_1}{2 \pi}
e^{i (-1)^{n} (s-s') y_1} = \delta(s-s')  \label{ortho} 
\eea 
since the integration over $x$ produces $2 \pi \delta(y_1+y_2)$ and after integration over $y_2$
almost all terms in the exponential cancel (i.e. for $y_2=-y_1$). \\

{\bf Explicit expressions}. 
Similar functions were defined in \cite{Penson} as Green functions of 
higher order diffusion equations, and some of their properties were studied there.
It was shown that they admit explicit expressions in terms of hypergeometric
functions. Here we give only the case $n=2$, where it reads (correcting for
some misprints in \cite{Penson})
\be
 \Ai_5(x) = \sum_{k=0}^3 c_k x^k \,  _0F_3(;\alpha_k;-\frac{x^5}{625})
\ee
where the $\alpha_k=(\frac{a_{k+1}}{5},\frac{a_{k+2}}{5},\frac{a_{k+3}}{5})$,
with the coefficients
$(a_1,..a_6)=(2,3,4,6,7,8)$ and the coefficients
$c_0 = 
\frac{2^{3/5} \sqrt{\pi }}{5^{4/5} \Gamma
   \left(\frac{3}{5}\right) \Gamma \left(\frac{7}{10}\right)}$, $c_1=
\frac{2 \sqrt[5]{2} \sqrt{\pi } }{5^{3/5} \Gamma
   \left(-\frac{1}{10}\right) \Gamma \left(\frac{6}{5}\right)}$, $c_2=
    \frac{\left(\sqrt{5}-1\right)  \Gamma \left(-\frac{2}{5}\right) }{2\ 5^{9/10}
   \sqrt{10-2 \sqrt{5}} \pi }$, $c_3=\frac{5^{3/10}  \Gamma
   \left(\frac{4}{5}\right)}{6 \sqrt{10-2
   \sqrt{5}} \pi }$.
These formula were used to plot the density $F_{1,4}$ in Fig. \ref{Fig_density}
in the text, and to calculate the Fredholm determinant giving the CDF of
$p_{\rm max}$, ${\cal F}_{4}$. \\
   
{\bf Large $|x|$ asymptotics}.
As for the standard Airy function the large argument asymptotics can be obtained 
by a saddle point calculation. We will perform it here only for $n=2$
\be
Ai_{5}(x) = \int_{- i \infty}^{+ i \infty} \frac{dz}{2 i \pi} e^{  - |x| z +  \frac{1}{5} z^{5} }
\ee

Let us start with $x>0$. The saddle points are solutions of
\bea
z_s^{4} = - x 
\eea
and they read
\bea
z_s = \pm (1 \pm i) \frac{1}{\sqrt{2}} x^{1/4} 
\eea 
This leads to
\bea
&& \Ai_{5}(x) \simeq e^{ - (1+i) \frac{2 \sqrt{2}}{5} x^{5/4}} \int_{-  \infty}^{+ \infty} \frac{dy}{2  \pi} 
e^{ - (1-i) \sqrt{2} x^{3/4} y^2} 
+ e^{ - (1-i) \frac{2 \sqrt{2}}{5} x^{5/4}} \int_{- \infty}^{+ \infty} \frac{dy}{2  \pi} 
e^{ - (1+i) \sqrt{2} x^{3/4} y^2} \\
&& = \frac{1}{2 \sqrt[4]{2} \sqrt{\pi } \sqrt{(1-i) x^{3/4}}} e^{ - (1+i) \frac{2 \sqrt{2}}{5} x^{5/4}}  
+ c.c
\eea  
where we have retained only the saddle point which lead to a decaying contribution.
This leads to, for $x \to +\infty$
\be
 \Ai_{5}(x) \simeq \frac{1}{\sqrt{2 \pi}} x^{-3/8} e^{- \frac{2 \sqrt{2}}{5} x^{5/4}} 
\cos( \frac{\pi}{8}-\frac{2 \sqrt{2}}{5} x^{5/4})
\ee
This formula allows to obtain the large $x>0$ asymptotics for the scaling function 
$F_{1,4}$ given in \eqref{righttail}. 

Let us study now $x<0$. The 4 saddle points are
\bea
z_s^4 = |x|  \Rightarrow z_s = \pm |x|^{1/4}  \quad \text{and} \quad z_s = \pm i |x|^{1/4}  
\eea 
Only the last two are relevant and they give for $x \to -\infty$
\bea \label{asympt2} 
&& \Ai_{5}(x) \simeq e^{ - i \frac{4}{5} |x|^{5/4}} \int_{-  \infty}^{+ \infty} \frac{dy}{2  \pi} 
e^{2 i |x|^{3/4} y^2} + cc  \\
&& = \frac{1}{2 \sqrt{2 \pi } \sqrt{-i |x|^{3/4}}} e^{ - i \frac{4}{5} |x|^{5/4}}  + c.c. 
 = 2 \sqrt{2} \frac{\cos \left(\frac{4 |x|^{5/4}}{5} - \frac{\pi}{4} \right)}{4 \sqrt{\pi } |x|^{3/8}}
\eea 
This form will now be used to show the matching of the edge density 
$F_{1,2}(s)$ with the density in the bulk. \\

{\bf Matching the density near the edge.} Using $F'_{1,4}(s) = - \Ai_{5}(s)^2$ one obtains for $s \to -\infty$, using \eqref{asympt2} 
\be
F'_{1,4}(s) \simeq -  \frac{\cos^2 \left(\frac{4 |s|^{5/4}}{5} - \frac{\pi}{4} \right)}{2 \pi |s|^{3/4}}
= -  \frac{1 + \sin \left(\frac{8 |s|^{5/4}}{5}  \right)}{4 \pi |s|^{3/4}} 
\ee 
Integrating one finds
\be\label{density_oscil}
F_{1,4}(s) \simeq \frac{1}{\pi} |s|^{1/4} - \frac{\cos \left(\frac{8 |s|^{5/4}}{5}  \right)}{8 \pi |s|} 
\ee
which, up to subleading oscillations, shows the result given in text below \eqref{Ain}.
\begin{figure}
\includegraphics[width = 0.5\linewidth]{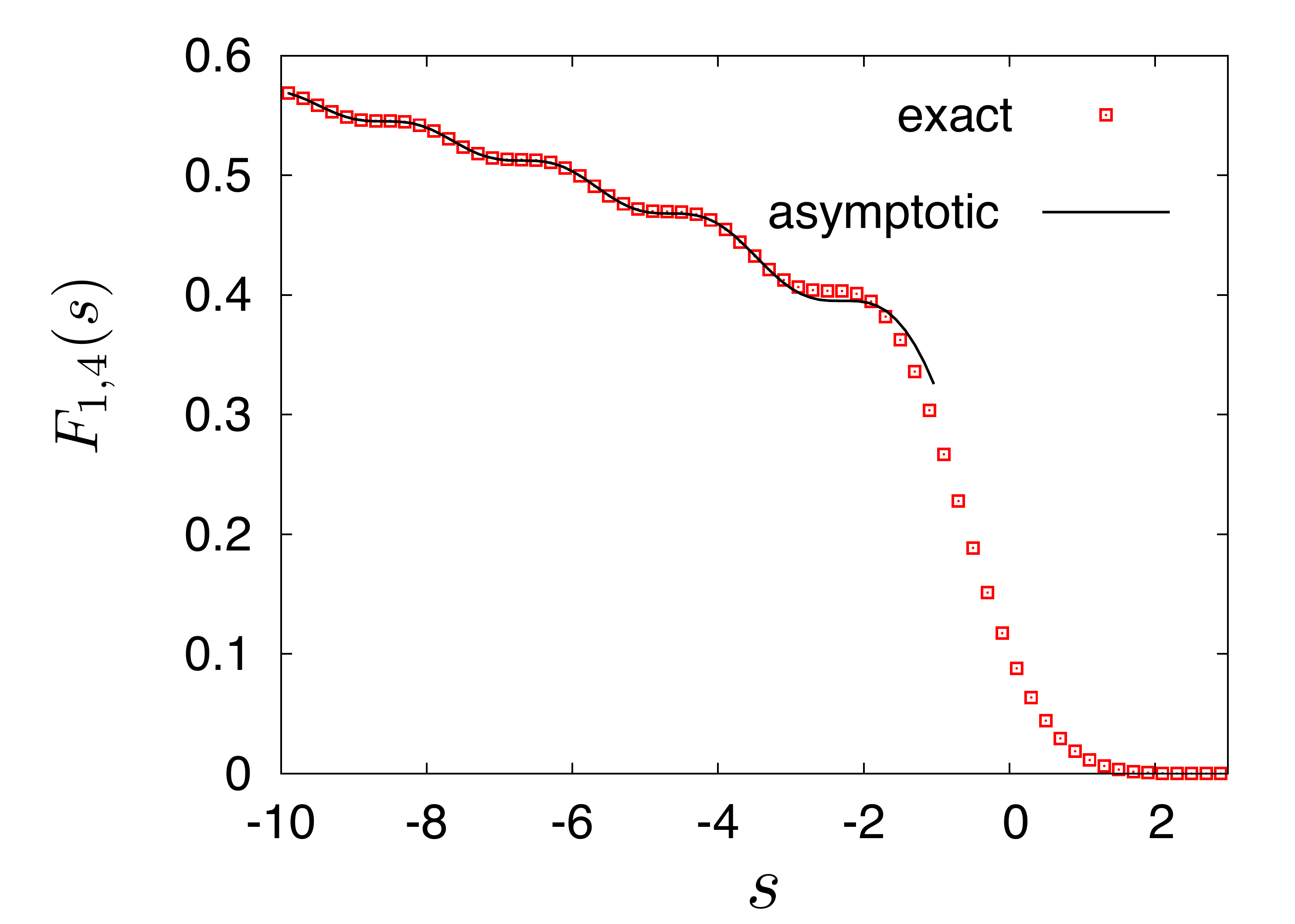}
\caption{Plot of the exact edge density $F_{1,4}(s)$ as given in Eq.~(\ref{densityedge}) (square symbols). The solid line corresponds 
to the asymptotic behaviour for $s \to - \infty$ given in Eq. (\ref{density_oscil}).}\label{compare_density}
\end{figure}
It establishes the smooth matching for the density between the negative $s$ side of the edge scaling
regime and the bulk regime. 
In Fig. (\ref{compare_density}) we show that this asymptotic behaviour (\ref{density_oscil}) describes
very accurately the oscillating behaviour of $F_{1,4}(s)$ for large negative $s$. Note finally that a similar matching with the bulk density 
is expected for all $n$ with
\be
F_{1,2n}(s) \simeq \frac{1}{\pi} |s|^{\frac{1}{2n}} \;.
\ee 
Of course, for arbitrary $n$, one also expects a (subleading) oscillating behaviour, as in Eq. (\ref{density_oscil}).

\medskip
\begin{center}
{\bf C. Derivation of the scaling form of the kernel near the edge}
\end{center}
\medskip

Let us give a simple argument to establish the scaling form of the kernel near the edge \eqref{Kscal}.
The energy eigenfunctions near the edge can be written more precisely as \eqref{PsiAin} 
\be \label{newpsi} 
 \hat \psi_k(p) \simeq c_k \Ai_{2 n +1} \left(\frac{p-p_e}{p_N} +\frac{m(\mu-\epsilon_k)}{p_e p_N} \right)\;. 
\ee 
where $c_k$ is an unknown coefficient which can be fixed e.g. using WKB approximation 
\cite{Eis2013}. Alternatively one can proceed as follows using the self-reproducibility property
of the kernel. Inserting \eqref{newpsi} into the kernel $K_\mu$, one finds that it takes the following
form for $p,p'$ near the edge
\be \label{sumK} 
K_\mu(p,p') \simeq \sum_k |c_k|^2  \Ai_{2 n +1} \left(\frac{p-p_e}{p_N} +\frac{m(\mu-\epsilon_k)}{p_e p_N} \right)  \Ai_{2 n +1} \left(\frac{p'-p_e}{p_N} +\frac{m(\mu-\epsilon_k)}{p_e p_N} \right)\;.
\ee
which can be checked to decay rapidly to zero if $|p-p' | \gg p_N$. In the large $\mu$ limit one can
replace the sums over the eigenstates by an integral and defining $u=\frac{m(\mu-\epsilon_k)}{p_e p_N}$, \eqref{sumK} takes
the form
\be \label{scalf} 
K_\mu(p,p') \simeq \int du \, c(u) \Ai_{2 n +1}(\frac{p-p_e}{p_N} + u) \Ai_{2 n +1}(\frac{p'-p_e}{p_N} + u) 
\ee
where $\sum_k |c_k|^2$ has been replaced by $\int du c(u)$ with an unknown positive function $c(u)$ related
to the coefficients $c_k$. Since $K_\mu(p,p')$ is self-reproducing one has
\be
\int dp'' K_\mu(p,p'') K_\mu(p'',p') = K_\mu(p,p')
\ee 
When $p,p'$ are both near the edge, the integral over $p''$ is dominated
by $p''$ close to the edge and one can thus use the scaling form \eqref{scalf}.
Using the orthonormality of the functions $\Ai_{2 n +1}$, see \eqref{ortho}, 
we see that it implies $p_N c(u)^2=c(u)$, hence $c(u)=1/p_N$
which demonstrates \eqref{Kscal}.

\medskip
\begin{center}
{\bf D. Some properties of the kernel $K_{\Ai,2 n+1}$ }
\end{center}
\medskip

Here we derive some useful properties of the kernel mentioned in the text.\\

{\bf Self-reproducing property.} As a consequence of the orthonormality of the $\Ai_{2n+1}$ functions, see 
\eqref{ortho}, 
the kernel 
\bea
K_{\Ai,2 n +1}(s,s')= \int_0^{+\infty} du \Ai_{2 n +1}(s+u) \Ai_{2 n +1}(s'+u) 
\eea 
is self-reproducing, since
\bea
 && \int_{-\infty}^{+\infty} ds'' K_{\Ai,2 n +1}(s,s'') K_{\Ai,2 n +1}(s'',s') \\
 && = \int_0^{+\infty} du_1 \int_0^{+\infty} du_2 \int_{-\infty}^{+\infty} ds'' 
 \Ai_{2n +1}(s+u_1) \Ai_{2n +1}(s''+u_1) \Ai_{2 n +1}(s''+u_2) \Ai_{2 n +1}(s'+u_2) = K_{\Ai,2 n +1}(s,s') \nn
\eea 
where we have used the orthonormality property \eqref{ortho} of the $\Ai_{2n+1}$ functions.\\


{\bf Differential equation for the kernel.}
From the definition of the kernel of sees that it satisfies the following differential equation
\bea
(\partial_{s}+\partial_{s'}) K_{\Ai,2 n +1}(s,s') = 
\int_0^{+\infty} du \partial_u [ \Ai_{2 n+1}(s+u) \Ai_{2 n +1}(s'+u) ] =  - \Ai_{2 n+1}(s) \Ai_{2 n +1}(s') \label{Kdiff} 
\eea
after integration since the function $\lim_{x \to +\infty} \Ai_{2 n+1}(x) = 0$. This generalises
the identity for the Airy kernel $n=1$. \\

{\bf Christoffel-Darboux type formula}. It is possible to express, for arbitrary $n$, the kernel directly in 
terms of the functions $\Ai_{2 n+1}$. Let us first note the following identity, for any smooth function $f$
and integer $n \geq 1$
\be
\partial_u  \sum_{p=0}^{n-1} (-1)^p \left( f^{(p)}(s+u) f^{(2 n-1-p)}(s'+u) - f^{(p)}(s'+u) f^{(2 n-1-p)}(s+u) \right)  = f(s+u) f^{(q)}(s'+u) - f(s'+u) f^{(q)}(s+u) 
\ee
Hence we have, using this identity,
\bea
&& (s-s') K_{\Ai,2 n +1}(s,s')= \int_0^{+\infty} du (s+u-(s'+u)) \Ai_{2 n+1}(s+u) \Ai_{2 n +1}(s'+u) \\
&& 
= (-1)^{n+1} \int_0^{+\infty} du \, \left[ \Ai_{2 n +1}^{(2 n)}(s+u) \Ai_{2n +1}(s'+u) - \Ai_{2 n +1}(s+u) \Ai_{2 n+1}^{(2 n)}(s'+u) \right] \nn \\
&& = (-1)^{n} \int_0^{+\infty} du 
 \partial_u  \sum_{p=0}^{n-1} (-1)^p \left( \Ai_{2 n +1}^{(p)}(s+u) \Ai_{2 n +1}^{(2n-1-p)}(s'+u) - \Ai_{2 n+1}^{(p)}(s'+u) \Ai_{2 n+1}^{(2 n-1-p)}(s+u) \right) \nn \\
 && = (-1)^{n-1} \sum_{p=0}^{n-1} (-1)^p ( \Ai_{2 n+1}^{(p)}(s) \Ai_{2n+1}^{(2n-1-p)}(s') - 
 \Ai_{2 n+1}^{(p)}(s') \Ai_{2 n+1}^{(2 n-1-p)}(s) ) \nn
\eea 
Finally we obtain
\bea
K_{\Ai,2 n +1}(s,s')= \frac{(-1)^{n-1} \sum_{p=0}^{n-1} (-1)^p ( \Ai_{2 n+1}^{(p)}(s) \Ai_{2n+1}^{(2n-1-p)}(s') - 
 \Ai_{2 n+1}^{(p)}(s') \Ai_{2 n+1}^{(2 n-1-p)}(s) )}{s-s'} \label{CD1} 
\eea 
For $n=1$ we recover the standard formula for the Airy kernel
\bea
K_{\Ai}(s,s')= \frac{1}{s-s'} [ \Ai(s) \Ai'(s') - \Ai'(s) \Ai(s') ]
\eea 
and for $n=2$ one obtains the formula \eqref{CD5} given in the text.

\medskip
\begin{center}
{\bf E. Finite temperature kernel}
\end{center}
\medskip

We give a short derivation of the finite $T$ form of the kernel given in the text, 
using the general method introduced in \cite{fermions_review}
which relates the finite $T$ kernel to the $T=0$ one.
For simplicity we set here $m=\hbar=1$ and restore the units at the end.
We consider the kernel in the grand canonical ensemble in momentum space,
denoted $K_{\tilde \mu,2n+1}$, at chemical
potential $\tilde \mu$ (see \cite{fermions_review} for definitions and details). 
We are interested in its form near the edge $p_e$. 
It can be obtained by inserting the zero temperature scaling form \eqref{Kscal} into Eq. 240 of \cite{fermions_review}, leading to 
\bea
K_{\tilde \mu,2n+1}(p,p') \simeq \int d\mu' \frac{1}{1 + e^{\beta (\mu' - \tilde \mu)} } \partial_{\mu'}
\frac{1}{p_N(\mu')} K_{\Ai,2n+1}( \frac{p-p_e(\mu')}{p_N(\mu')} , \frac{p'-p_e(\mu')}{p_N(\mu')})  
\eea 
where $p,p'$ are near the edge, $p_e(\mu')=\sqrt{2 \mu'}$ and $p_N(\mu')=(g/p_e(\mu'))^{1/(2n+1)}$. We
recall that near the edge
$\tilde \mu \approx \mu$ \cite{fermions_review}. Note that one can neglect the
action of $\partial_\mu$ on the factor $1/p_N(\mu)$.
One has
\bea
&& \frac{p-p_e(\mu')}{p_N(\mu')} = \frac{p-p_e(\mu)}{w_N( \mu)}
-  \beta_\mu (\mu'-\mu) + O((\mu'-\mu)^2)  \\
&& \beta_\mu =  \frac{1}{p_N(\mu)} \partial_\mu p_e(\mu) = 
g^{-1/(2n+1)} p_e^{1/(2n+1)} \partial_\mu p_e(\mu) = g^{-1/(2n+1)}  2^{-\frac{n}{2n+1}} \mu^{-\frac{n}{1+2n}} 
\eea 
the first term is subdominant. We define $\beta (\mu'-\mu) = - b u$, hence
\bea
&& \frac{p-p_e(\mu')}{p_N(\mu')} = \frac{p-p_e(\mu)}{p_N( \mu)}
+ b  \frac{\beta_\mu}{\beta} u  + O((\mu'-\mu)^2) 
\eea
This leads to the definition of the scaled inverse temperature 
$b$ as given in the text, Eq. \eqref{defb} (restoring units), 
so that $b  \frac{\beta_\mu}{\beta}  = 1$. Hence we find that the kernel takes the scaling form
\bea
&& K_{\mu,2n+1}(p,p') \simeq - \frac{1}{p_N(\mu)}  
\int du \frac{1}{1 + e^{- b u} } \partial_{u}
K_{{\rm Ai},2n+1}( \frac{p-p_e(\mu)}{p_N(\mu)} + u , \frac{p'-p_e(\mu)}{p_N(\mu) + u })  
\\
&& =  \frac{1}{p_N(\mu)} \int_{-\infty}^{+\infty} du \frac{1}{1 + e^{- b u} } 
{\rm Ai}_{2n+1}(\frac{p-p_e(\mu)}{p_N(\mu)} + u) {\rm Ai}_{2n+1}(\frac{p'-p_e(\mu)}{p_N(\mu)} + u) 
\\
&& = \frac{1}{p_N(\mu)} K_{2n+1,b}(\frac{p-p_e(\mu)}{p_N(\mu)} , \frac{p'-p_e(\mu)}{p_N(\mu)}) 
\eea 
where $K_{2n+1,b}$ is given in the text in \eqref{kernelT}.\\

{\bf High temperature limit.} Let us examine the high temperature limit of the edge regime, i.e. 
small $b = T^*/T \ll 1$. The following property is shown in \cite{AlexST}.
Consider any determinantal process $k_i \in \mathbb{R}$ characterised by a kernel $K(k,k')$ of the form 
\bea
K(k,k') = \int du \frac{1}{1 + e^{- b u}} f(k+u) f(k'+u) 
\eea 
with a density $\rho_K(k)=K(k,k)$ which behaves at large negative $k$ as $\rho_K(k) \simeq_{k \to -\infty} A |k|^{\alpha}$.  
Then, it is shown in \cite{AlexST} that, in the limit of small $b$, 
the CDF of the variable $\xi= b k$ 
takes the large deviation form
\bea
&& {\rm Prob}(\xi< r) \sim \exp(- b^{- \frac{1+\alpha}{\alpha} } \Psi(e^{- r}) ) \quad , \quad  \Psi(z) := A \int_{-\infty}^0 da  |a|^{\alpha}  \ln(1+ z e^{a}) 
\eea

For the present application we will set $\alpha=1/(2n)$, $A=1/\pi$ and $K=K_{\Ai, 2n+1}$,
$f=\Ai_{2n+1}$, and consider the determinantal process $k_i=\frac{p_i(T)-p_e}{p_N}$,
where the $p_i$ are the momenta of the fermions in the grand canonical ensemble at temperature $T$.
Introducing the scaled variable
\be
\xi = b k = b \frac{p_i(T)-p_e}{p_N}
\ee 
we find, from the above result, that for small $b \ll 1$
\bea
&& {\rm Prob}(\xi<r) = \exp(- b^{- \frac{2n+1}{2 n} } \Psi_{2n}(e^{- r}) ) \quad , \quad  \Psi_{2n}(z) = - \frac{1}{\pi} \Gamma(1+ \frac{1}{2n}) {\rm Li}_{2+ \frac{1}{2 n}}(-z) 
\eea
This is a generalisation, for arbitrary $n$, of the result of 
\cite{UsShortTime} for $n=1$. 
Hence in the regime of typical fluctuations, expanding $\Psi_{2n}(z)$ to $O(z)$, we find that
\bea
 {\rm Prob}(\xi<r) = \exp(- b^{- \frac{2n+1}{2 n} } \frac{1}{\pi} \Gamma(1+ \frac{1}{2n}) e^{- r} )
\eea 
Hence the typical value of $\xi$ is $\xi_{typ}=\ln( b^{- \frac{2n+1}{2 n} } \frac{1}{\pi} \Gamma(1+ \frac{1}{2n}) )$, and the typical fluctuations are of Gumbel type, i.e. one has $\xi=\xi_{typ} + G$
where $G$ is a unit Gumbel random variable. In particular the variance is
\bea
{\rm Var}[ b \frac{p_{\rm max}(T) -p_e}{p_N} ]  = \frac{\pi^2}{6} 
\eea 
as given in the text, below Eq. \eqref{kernelT}.

\medskip
\begin{center}
{\bf F. Painlev\'e II hierarchy and asymptotics of the Fredholm determinant ${\cal F}_{2n}$}
\end{center}
\medskip

{\bf Painlev\'e II hierarchy.}
Let us recall here the definition of the Painlev\'e II hierarchy of non linear differential equations.
We follow e.g. Eqs. (1.31-1.32) of Ref. \cite{Claes1} (and see references therein). 
The $n$-th member of the Painlev\'e II hierarchy is a differential equation for
the function $g(x)$, which reads (for the special case $\tau_i=\alpha_i=0$ which is relevant here) 
\be \label{hierarchy1} 
 (\partial_x + 2 g) L_n[g' - g^2] = x g 
\ee
where the $L_j$ are operators which transform a function into a function and are
defined recursively as
\be
\partial_x L_{j+1} f = (\partial_x^3 + 4 f \partial_x + 2 f') L_{j} f  \quad , \quad L_{0} f = \frac{1}{2} 
\ee
For $n=1$, one finds $L_{1} f = f$ and the standard Painlev\'e II equation
\be
g''(x) = x g(x) + 2 g(x)^3  \label{PII1g} 
\ee
given in the text for the function denoted there $q(s)=g(s)$.
Next, for $n=2$ one finds $L_{2} f = f'' + 3 f^2$ and
\be
g''''(x) = x g(x) + 10 g(x) g'(x)^2 + 10 g(x)^2 g''(x) - 6 g(x)^5 \label{PII2g} 
\ee
which leads to \eqref{PII2new} in the text, for the function $q(s)=g(-s)$.
More generally the function $q(s)$ defined in the text for arbitrary $n$ is
given by $q(s)=g((-1)^{n+1} s)$. \\

{\bf Asymptotics of the Fredholm determinant ${\cal F}_{2n}$.}
Let us study the behaviour of $q(s)=g((-1)^{n+1} s)$ for $s \to -\infty$. Examination
shows that in the equations \eqref{PII1g}
and \eqref{PII2g} one can neglect all derivatives, to leading order for large negative $s$ (with 
$x=(-1)^{n+1} s$). Going back
to the recursion relation of the PII hierarchy \eqref{hierarchy1} these simplified equations can 
be obtained more systematically by performing the same approximation, leading to the simplified
recursion 
\bea
&& 2 g L_n[- g^2] = x g \quad , \quad \partial_x L_{j+1} f = (4 f \partial_x + 2 f') L_{j} f  \quad , \quad L_{0} f = \frac{1}{2} 
\eea 
The second equation is solved by writing $L_j f=a_j f^j$ and leads to the recursion
relation $(j+1) a_{j+1} = 4 (j + \frac{1}{2}) a_j$ with $a_0=\frac{1}{2}$, hence
\be
a_j = 2^{2 j-1} \frac{\Gamma(j + \frac{1}{2})}{\sqrt{\pi} \Gamma(j+1)} 
\ee 
with $a_0=1/2$, $a_1=1$, $a_2=3$. It finally leads to
\bea \label{asymptq}
q(s) \simeq \kappa_n (-s)^{\frac{1}{2 n}}  \quad , \quad \kappa_n = 
\frac{1}{2} [ \frac{\sqrt{\pi} \Gamma(n+1)}{\Gamma(n + \frac{1}{2})} ]^{\frac{1}{2 n}} 
\eea 

We now recall that the Fredholm determinant \eqref{calF} is obtained from
the function $q(s)$, solution of the Painlev\'e equation $P_{\rm II}^{(n)}$ with the
boundary condition $q(s) \sim_{s \to +\infty} \Ai_{2 n+1}(s)$, 
through the formula \eqref{F2nint} (see derivation in Section G)
\bea
\ln {\cal F}_2(s) = - \int_{s}^{+\infty} du (u-s) q(u)^2 
\eea 
Plugging in \eqref{asymptq}, this leads to the result given in the text below Eq. \eqref{PII2new}
for $s \to -\infty$
\bea
\ln {\cal F}_{2n}(s) \simeq - f_n (-s)^{2 + \frac{1}{n}}  \quad , \quad f_n = 
\frac{n^2}{4(2 n+1)(n+1)} [ \frac{\sqrt{\pi} \Gamma(n+1)}{\Gamma(n + \frac{1}{2})} ]^{\frac{1}{n}} 
\eea 
For $n=2$ we find, explicitly
\bea
q(s) \simeq \frac{1}{6^{1/4}} |s|^{1/4} 
-\frac{5}{16\ 6^{3/4} |s|^{9/4}}-\frac{99\
   3^{3/4}}{512 \sqrt[4]{2} |s|^{19/4}} + .. \quad , \quad \ln {\cal F}_4(s) \simeq - \frac{2 \sqrt{2}}{15 \sqrt{3}} (-s)^{5/2} + O(s) 
\eea 
This asymptotics for ${\cal F}_4(s)$ is plotted in Fig. \ref{Fig_density}.

\bigskip
\begin{center}
{\bf G. Differential equation for the Fredholm determinant ${\cal F}_{2n}$ and the 
connection to the second Painlev\'e hierarchy}
\end{center}
\medskip

In this section we obtain the differential equation satisfied by the Fredholm
determinant ${\cal F}_{2n}$ in Eq. \eqref{calF} of the text. To this aim 
we first define the problem in a more general framework. We consider 
the following Fredholm determinant
\bea
F(s) := {\rm Det}(I- \hat K) \quad , \quad \hat K = K P_s \Leftrightarrow \hat K(x,y) = K(x,y) \theta(y-s) \label{Fs} 
\eea 
where $K$ is a kernel which satisfies the following three properties:
\begin{itemize}
\item {\bf property 1}: The kernel can be constructed from a function $\phi(x)$ and its derivatives as
\bea
K(x,y) = \langle x | K |y \rangle = \frac{1}{x-y} 
\left( (-1)^{n-1} \sum_{p=0}^{n-1} (-1)^p ( \langle x | \phi^{(p)} \rangle \langle \phi^{(2n-1-p)}|y \rangle 
- \langle x | \phi^{(2n-1-p)} \rangle \langle \phi^{(p)}|y \rangle
 \right) \label{prop1}
\eea
in quantum mechanical notations, 
where $|\phi\rangle=|\phi^{(0)}\rangle$ is the ket (vector) associated to the function
$\langle x|\phi\rangle=\phi(x)$ and $|\phi^{(p)}\rangle$ denote the kets associated to the 
derivatives $\langle x|\phi^{(p)}\rangle=\phi^{(p)}(x)$.

\item {\bf property 2}: The kernel satisfies 
\be
(\partial_x + \partial_y) K(x,y) = - \phi(x)\phi(y) \label{prop2}
\ee

\item {\bf property 3}: The function $\phi(x)$ obeys the differential equation
\bea
\phi^{(2n)}(x) = (-1)^{n+1}  \phi(x) \label{prop3}
\eea 

\end{itemize}

One recognises that $\phi(x)=\Ai_{2n+1}(x)$ and $K=K_{\Ai,2 n+1}$ defined in the text satisfy these
three properties, with $F(s)= {\cal F}_{2n}(s)$.
The first is the Christoffel-Darboux property \eqref{CD1}, the second is \eqref{Kdiff}
and the third is \eqref{Aidiff}. Whenever these three properties hold, a closed system of differential equations can be derived for the Fredholm determinant. Here we follow and extend the derivation of Tracy and Widom
in \cite{TracyWidomAiry} who treated the case $n=1$ (Airy kernel). A similar, but somewhat different, extension was obtained by Br\'ezin and Hikami in \cite{BrezinHikamiSource}, to which we also refer. 
We first derive the equations, and in a second part we analyse them and relate them to the second
Painlev\'e hierarchy. \\

{\bf Derivation of the differential equations.} 
We start by applying $\partial_s$ to the logarithm of \eqref{Fs} which leads to
\bea  \label{derFred} 
\partial_s \ln F(s) = \partial_s {\rm Tr} \ln (I-\hat K) = - {\rm Tr} (1-\hat K)^{-1}  \partial_s \hat K 
=  {\rm Tr} (1-\hat K)^{-1}  K |s\rangle \langle s|
= \tilde K(s,s) \quad , \quad \tilde K = \frac{\hat K}{1- \hat K} 
\eea 
which is simply the diagonal element of the operator $\tilde K$. We have used the expression for the 
derivative
\bea
\partial_s \hat K(x,y) = - K(x,y) \delta(y-s) \Leftrightarrow \partial_s \hat K = - K |s\rangle \langle s|
\eea 
where we recall that $\langle x |y \rangle=\delta(x-y)$. We now derive a differential equation
satisfied by $\tilde K(s,s)$. The route is an extension of \cite{TracyWidomAiry} and requires
introducing two sets of $2 n$ auxiliary functions. Before doing so, let us define the two operators,
the position $X$, and the derivative $D$, as follows. For any $x,y$ and ket vector $|f \rangle$
\bea
\langle x|D|f \rangle=f'(x)=-\langle f|D|x \rangle \quad , \quad \langle x |X  =  x \langle x |   \quad , \quad 
X |y \rangle =y | y \rangle
\eea
where $\langle x|f\rangle=f(x)$. To manipulate them, we need to recall the 
operator commutator and derivation identities \cite{TracyWidomAiry} 
\bea
[L , (1-\hat K)^{-1}] = (1-\hat K)^{-1} [L,\hat K] (1-\hat K)^{-1} \quad , \quad 
\partial (1-\hat K)^{-1} =  (1-\hat K)^{-1} \partial \hat K (1-\hat K)^{-1} \label{id2}
\eea 
which we will use repeatedly for $L=D$ or $L=X$. \\

Let us now use {\bf property 1}, i.e. Eq. \eqref{prop1}. Since $ \langle x [X, K] |y \rangle= (x-y) K(x,y)$ we obtain, upon further right multiplication 
by $P_s$ (using that $X$ and $P_s$ commute)
\bea
[X, \hat K] = (-1)^{n-1} \sum_{p=0}^{n-1} (-1)^p ( |\phi^{(p)} \rangle \langle \phi^{(2n -1-p)} | - 
|\phi^{(2n -1-p)} \rangle \langle \phi^{(p)} |) P_s
\eea
Hence, using that $\tilde K=(I- \hat K)^{-1}-I$, 
\bea
&& (x-y) \tilde K(x,y) = 
\langle x|[X, \tilde K] |y\rangle 
= \langle x|  (1-\hat K)^{-1} [X,\hat K] (1-\hat K)^{-1}  |y\rangle \\
&& =  (-1)^{n-1}  \sum_{p=0}^{n-1}  (-1)^p
\langle x|  (1-\hat K)^{-1}  ( |\phi^{(p)} \rangle \langle \phi^{(2 n -1-p)} | - 
|\phi^{(2 n-1-p)} \rangle \langle \phi^{(p)} |) P_s (1-\hat K)^{-1}  |y\rangle \\
&& = (-1)^{n-1}  \sum_{p=0}^{n-1}  (-1)^p [ 
q_p(s;x) \tilde q_{2 n-1-p}(s;y) - q_{2 n-1-p}(s;x)  \tilde q_{p}(s;y) ]
\eea 
where we have defined the $2n$ auxiliary functions, for $p=0,.. 2n-1$
\bea
&& q_p(s;x) := \langle x| (1-\hat K)^{-1} | \phi^{(p)} \rangle
= \phi^{(p)}(x) + \int_{y>s} K(x,y) \phi^{(p)}(y) + \int_{y_1>s,y_2>s} K(x,y_1) K(y_1,y_2) \phi^{(p)}(y_2) 
+ \cdots \\
&& \tilde q_p(s;x) : = \langle \phi^{(p)} | P_s (1-\hat K)^{-1} |  x \rangle  
= \theta(y-s) [ \phi^{(p)}(y) + \int_{x>s} \phi^{(p)}(x) K(x,y)  + \int_{x_1>s,x_2>s} \phi^{(p)}(x_1) 
K(x_1,x_2) K(x_2,y) 
+ \cdots ] \nn
\eea 
Note that $q_p(s;x)$ is smooth, while $\tilde q_p(s;x)$
is not smooth (it vanishes for $x<s$), and in fact one has $\tilde q_p(s;y) = q_p(s;y) \theta(y-s)$.
Hence in the following we only need $n$ auxiliary functions $q_p$.
This leads to (for $y \to s^+$, the same limit procedure as in \cite{TracyWidomAiry}) 
\bea
\tilde K(s,s) = (-1)^{n-1}  \sum_{p=0}^{n-1}  (-1)^p [ 
\partial_x q_p(s;x)|_{x=s} q_{2 n -1-p}(s;s) - \partial_x q_{2n -1-p}(s;x)|_{x=s} q_{p}(s;s) ] \label{Ktilde1} 
\eea 

We now establish a set of differential equations for the functions $q_p(s;x)$. It then leads to establishing 
two closed sets of $2 n$ differential equations (each) for $q_p(s)$ and $u_p(s)$, respectively, $p=0,\cdots 2n-1$, defined as
\bea
q_p(s) := q_p(s;s)= \langle x| (1-\hat K)^{-1} | \phi^{(p)} \rangle \quad , \quad u_p(s) :=  \langle \phi | P_s (1-\hat K)^{-1} | \phi^{(p)} \rangle \label{defu} 
\eea 

The first set of differential equations reads (in two parts)
\bea
&& \partial_x q_p(s;x) = q_{p+1}(s;x) - q_0(s;x) u_p(s)  + \tilde K(x,s)  q_p(s) \quad , \quad \partial_s q_p(s) = q_{p+1}(s) - q_0(s) u_p(s) \label{eqs1} 
\eea
To obtain it we use {\bf property 2}, i.e. Eq. \eqref{prop2}. It allows to establish that 
\bea
[D, \hat K] = - |\phi \rangle \langle \phi | P_s + K | s \rangle  \langle s| \label{prop11} 
\eea
Indeed one has
\bea
&& \langle x|[D, \hat K] |y\rangle = \partial_x( K(x,y) \theta(y-s) ) + \partial_y ( K(x,y) \theta(y-s) )
= [ (\partial_x + \partial_y) K(x,y) ] \theta(y-s) + \langle x |K | s \rangle  \langle s| y \rangle 
\eea 
Using {\bf property 2}, i.e. Eq. \eqref{prop2}, this leads to \eqref{prop11}.\\

Using \eqref{id2} and \eqref{prop11} we now calculate the derivative
\bea \label{ddx} 
&& \partial_x q_p(s;x) = \langle x| D (1-\hat K)^{-1} | \phi^{(p)} \rangle = q_{p+1}(s;x) + \langle x| [D , (1-\hat K)^{-1}] | \phi^{(p)} \rangle  \\
&& = q_{p+1}(s;x) + \langle x| (1-\hat K)^{-1} [D, \hat K] (1-\hat K)^{-1} | \phi^{(p)} \rangle 
=  q_{p+1}(s;x) + \langle x| (1-\hat K)^{-1} [- |\phi \rangle \langle \phi | P_s + K | s \rangle  \langle s| ] 
(1-\hat K)^{-1} | \phi^{(p)} \rangle \nn \\
&& = q_{p+1}(s;x) - q_0(s;x) \langle \phi | P_s (1-\hat K)^{-1} | \phi^{(p)} \rangle
+ \tilde K(x,s)  q_p(s;s) \nn
\eea
which establishes, using the definition \eqref{defu}, the first part of \eqref{eqs1}. We have used that $P_s |s\rangle = |s\rangle$ which 
implies that
$\tilde K(x,s)=\langle x|(1- \hat K)^{-1} K |s \rangle=\langle x|(1- \hat K)^{-1} \hat K |s \rangle$.
The second part of \eqref{eqs1} is established in the same way. We calculate the total
derivative with respect to $s$
\bea
&& \partial_s q_p(s;s) = \langle s| D (1-\hat K)^{-1} | \phi^{(p)} \rangle   +
\langle s|  (1-\hat K)^{-1} \partial_s \hat K (1-\hat K)^{-1} | \phi^{(p)}  \rangle 
\eea
Hence it is similar to \eqref{ddx} except that there is an additional term
\bea
&& 
\langle s|  (1-\hat K)^{-1} \partial_s \hat K (1-\hat K)^{-1} | \phi^{(p)}  \rangle 
=  - \langle s|  (1-\hat K)^{-1} K |s\rangle \langle s| (1-\hat K)^{-1} | \phi^{(p)}  \rangle
= - \tilde K(s,s)  q_p(s;s)
\eea
which simply 
cancels the term $\tilde K(s,s)  q_p(s)$ in the first part of \eqref{eqs1} 
and leads to the second part of \eqref{eqs1}.

We now establish the differential equations for the second set of auxiliary functions, $u_p(s)$. We first 
calculate 
\bea
&& \partial_s [ P_s (1-\hat K)^{-1}] 
= - ( |s \rangle + P_s (1-\hat K)^{-1} K |s \rangle ) \langle s| (1-\hat K)^{-1}
\\
&& = - ( P_s + P_s (1-\hat K)^{-1} \hat K |s \rangle ) \langle s| (1-\hat K)^{-1} 
= - P_s (1-\hat K)^{-1}   |s \rangle  \langle s| (1-\hat K)^{-1} 
\eea 
where we used again that $P_s |s \rangle = |s \rangle$. Hence
\bea
&& \partial_s u_p(s) = \langle \phi | (\partial_s [ P_s (1-\hat K)^{-1}])  | \phi^{(p)} \rangle
= - \langle \phi | P_s (1-\hat K)^{-1}   |s \rangle  \langle s| (1-\hat K)^{-1}   | \phi^{(p)} \rangle \\
&& 
= - \tilde q_0(s,s) q_p(s) = - q_0(s) q_p(s)
\eea 
since $\tilde q_0(s,s)=\tilde q_0(s^+,s)=q_0(s)$.

Since \eqref{eqs1} relates $q_p$ and $q_{p+1}$ we still need to close the equations.
Fortunately one can now use the {\bf property 3}, i.e. Eq. \eqref{prop3}, which can also
be written as $|\phi^{(2n)} \rangle = (-1)^{n+1} X |\phi \rangle$. It allows 
to express $q_{2n}$ in terms of the $q_p$, and $u_p$, for $p=0,..2n-1$, hence it closes 
the equations. One has
\bea
&& q_{2 n}(s;x) = \langle x| (1-\hat K)^{-1} | \phi^{(2n)} \rangle = (-1)^{n+1}
\langle x| (1-\hat K)^{-1} X | \phi \rangle
= (-1)^{n+1} ( x q_0(s;x) + \langle x| [(1-\hat K)^{-1}, X] | \phi \rangle ) \\
&& = (-1)^{n+1} ( x q_0(s;x) - \langle x| (1-\hat K)^{-1} [X, \hat K] (1-\hat K)^{-1} | \phi \rangle ) \\
&& =  (-1)^{n+1} x q_0(s;x)  - \big(
\langle x| (1-\hat K)^{-1}  [ \sum_{p=0}^{n-1} (-1)^p ( |\phi^{(p)} \rangle \langle \phi^{(2 n-1-p)} | - 
|\phi^{(2 n-1-p)} \rangle \langle \phi^{(p)} |) ] P_s
(1-\hat K)^{-1} | \phi \rangle \big) \\
&& = (-1)^{n+1} x q_0(s;x) -
\sum_{p=0}^{n-1} (-1)^p ( q_p(s;x) u_{2 n-1-p}(s) - q_{2 n -1-p}(s;x) u_{p}(s) ) 
\eea 
We use this equation for $x=s$. We have used that 
$P_s (1-\hat K)^{-1} = P_s + P_s K P_s + P_s K P_s K P_s + \cdots$ is a
symmetric operator, hence $\langle \phi^{(p)} | P_s (1-\hat K)^{-1} | \phi \rangle =
 \langle \phi | P_s (1-\hat K)^{-1} | \phi^{(p)} \rangle =u_p(s)$.
 
Finally, we can go back to the equation \eqref{Ktilde1} and perform
the derivatives using the equations \eqref{eqs1}. It leads to 
\bea
\tilde K(s,s) = (-1)^{n-1}  \sum_{p=0}^{n-1}  (-1)^p [ 
q'_p(s) q_{q-1-p}(s) - q'_{q-1-p}(s) q_{p}(s) ]
\eea\\

{\bf Summary of the equations.}
Let us now recapitulate the system of equations obtained above for
any fixed integer $n$. 
The $4n$ functions $\{ q_p(s),u_p(s) \}_{0 \leq p \leq 2 n-1}$ satisfy
the closed system
\bea \label{system1} 
&& q_p' = q_{p+1} - q_0 u_p  \quad , \quad u_p' = - q_0 q_p \quad , \quad 0 \leq p \leq 2 n -1
\eea
where, in addition
\bea \label{system2} 
q_{2 n}  = (-1)^{n+1} s q_0 -
\sum_{p=0}^{n-1} (-1)^p ( q_p u_{2 n -1-p} - q_{2 n -1-p} u_{p} ) 
\eea
From their definitions $\{ q_p(s),u_p(s) \}_{0 \leq p \leq 2 n-1}$
in \eqref{defu}, they vanish for $s \to +\infty$. This system 
allows to calculate $\tilde K(s,s)$ from 
\bea \label{Kss2} 
\tilde K(s,s) = (-1)^{n-1}  \sum_{p=0}^{n-1}  (-1)^p [ 
q'_p(s) q_{q-1-p}(s) - q'_{q-1-p}(s) q_{p}(s) ]
\eea
Remarkably, using the above equations one can check that
\bea
\partial_s \tilde K(s,s) = - q_0(s)^2  \label{derK} 
\eea 
Hence, to summarise, the Fredholm determinant
satisfies 
\bea
\partial_s^2 \ln F(s) = - q_0(s)^2 \label{miracle} 
\eea 
where $q_0(s)$ is determined from the system \eqref{system1}, \eqref{system2}. 
Note that this equation \eqref{miracle} corresponds to \eqref{F2nint} in the
text, where we denote $q_0(s)=q(s)$. 

For $n=1$ it is interesting to compare with the results of Tracy and Widom.
One recovers the equations (1.6-1.9) of \cite{TracyWidomAiry}: 
one must consider $j=1$ there, i.e. remove the
index $j$ everywhere and discard the terms $\sum_{k \neq j}$, with 
$a_j=a=s$, $u_0=u$, $u_1=v$, $q_0=q$ and $q_1=p$. 
Eq. \eqref{system2} here is then the same as (1.10) there,
and Eq. \eqref{Kss2} the same as (2.8-2.10) there.
Finally \eqref{miracle} is (1.14) there. \\

{\bf Analysis of the equations: Painlev\'e II hierarchy.}
We now analyse the equations \eqref{system1}, \eqref{system2}, \eqref{miracle}
and establish the connection to the Painlev\'e II hierarchy, as discussed in the text.
We find that the function $q_0(s)$, which is denoted $q(s)$ in the text,
satisfies a {\it closed} differential equation, which coincides 
with the $n$-th member of the Painlev\'e II hierarchy. 
The latter is recalled in section F.

To analyse the flow, it is useful to identify invariants. 
We find that there are $n$ quadratic invariants $I_p$, $1 \leq p \leq n$, 
\bea
I_p = u_{2 p-1} + \frac{1}{2} \sum_{k=0}^{2 p-2} (-1)^{k+1} (u_k u_{2p-2-k} - q_k q_{2p-2-k} ) = 0 
\eea 
Note that they do not depend explicitly on $n$, i.e. the invariants for $n'<n$ are a subset of those for $n$.
It is immediate to check that
\bea
\partial_s I_p = - q_0 q_{2p-1} - \frac{1}{2} \sum_{k=0}^{2 p-2} (-1)^{k+1}  (q_{k+1} q_{2 p - 2-k} 
+ q_{k} q_{2 p - 1-k} ) = 0
\eea 
using that the sum is telescopic of the form $\sum_{k=0}^{2 p-2} (-1)^{k+1}  (Q_k + Q_{k+1})=- Q_0 - Q_{2p-1}$ with $Q_k =q_{k} q_{2 p - 1-k}$. Furthermore since all the functions vanish at $s \to +\infty$
along the flow the invariants exactly vanish $I_p(s)=0$.

Remarkably, using these invariants, it is possible to eliminate all functions apart from $q_0$, and write a closed equation
for $q_0$. Using the condition that the invariants vanish, $I_p=0$, $p=1,..n$, we express the odd
$u_{2k+1}(s)$ as a function of all other functions. Then it is possible to obtain recursively all $q_p$
for $1 \leq p \leq 2 n - 2$ only in terms of $q_0$ and its derivatives and of the even
$u_{2k}(s)$ (not their derivatives). Finally one calculates the $2n$-th derivative of 
$q_0$, $q_0^{(2 n)}$, using \eqref{system1} and \eqref{system2} as well as these substitutions. 
All dependence in the $u_{2k}(s)$ then cancels. 

For $n=1$, from \eqref{system1} we have $q''_0=q_1'-(q_0 u_0)'= q_1'- q_0' u_0 + q_0^3$, and
using $q_1'= q_2 - q_0 u_1=
s q_0 + q_1 u_0 - 2 q_0 u_1$ from \eqref{system2}, we obtain 
$q_0''= s q_0 + q_1 u_0 - 2 q_0 u_1 - q_0' u_0 + q_0^3$.
Using that $q_1=q_0'+q_0 u_0$ and replacing $u_1$ using the invariant $I_1=u_1 + \frac{1}{2} (u_0^2-q_0^2)=0$ we see that $u_0$ cancels and we are left with 
 $q_0''= s q_0 + 2 q_0^3$, the usual Painlev\'e II equation. 

For $n=2$ the same procedure yields, after a tedious calculation, the following equation for
$q_0$ (as given in the text, Eq. \eqref{PII2new} with $q_0=q$)
\bea
q_0'''' = 
q_0 \left(10 (q_0')^2+10 q_0 q_0''-6
   q_0^4-s\right) \label{PII2} 
\eea
which is exactly (up to a change of sign of the argument) 
$P_{\rm II}^{(2)}$ a particular case of the second member of the second Painlev\'e
hierarchy, as discussed in Section F. 

We now conjecture, as announced in the text, that this property, i.e. $q_0(s) = g((-1)^{n+1} s)$ 
where $g(x)$ is the solution of the $n$-th member of the second Painlev\'e
hierarchy, recalled in Section F, hold for arbitrary $n$. 
We have checked this conjecture with Mathematica, using
the above differential system, up to large values of $n$.\\


{\bf Hamiltonian structure of the differential system.}
It is interesting to point out another remarkable property of the above system of differential 
equations \eqref{system1}, \eqref{system2}. Indeed, for $n=1$ it is known that there is a Hamiltonian
structure underlying the Painlev\'e II equation, see \cite{TracyWidomAiry} and references therein. 
Here also, we find a Hamiltonian structure, which is quite simple. The Hamiltonian reads
\bea
H[q,u] = (-1)^n \frac{1}{2} s q_0^2 + \frac{1}{2} \sum_{p=1}^{2 n -1} (-1)^{p+1}  q_p q_{2 n - p} 
+ q_0  \sum_{p=0}^{2 n -1} (-1)^{p+1}  u_p q_{2 n -1 - p}  
\eea 
Then one can check that the equations \eqref{system1}, \eqref{system2} can be written as
\bea
q'_p = (-1)^p \frac{\partial H[q,u]}{\partial q_{2 n -1 - p}}   \quad , \quad u'_p = (-1)^{p+1} 
\frac{\partial H[q,u]}{\partial u_{2 n -1 - p}}   \quad , \quad p=0,\cdots,2n-1
\eea 
Note that
\bea
\frac{d H}{ds} = \frac{\partial H}{\partial s} = \frac{1}{2} q_0^2 = - \frac{1}{2} u_0'
\eea 
Hence $\tilde H = H + \frac{u_0}{2}$ is an additional invariant of motion, which is cubic.
Comparing with \eqref{derK} we see that 
\bea
- H[q(s),u(s)] = \tilde K(s,s) 
\eea 
since both vanish for $s \to +\infty$, related to the derivative of the logarithm of the Fredholm 
determinant ${\cal F}_{2n}$ through Eq. \eqref{derFred}. Finally we note that for the Painlev\'e II hierarchy there exists a Hamiltonian formulation \cite{Mo-Hamiltonian}, which however looks much more complicated.
A route to prove our conjecture would be to establish the equivalence between the two Hamiltonian
dynamics. It is worth pointing out that, as a byproduct of our work, we have found a simple
Hamiltonian structure for arbitrary $n$.
%
%
%

\end{widetext}


\begin{thebibliography}{}
\bibitem{footnote0} {We denote this width $p_N$ for convenience, not to be confused
with $p_{i=N}$.}
\bibitem{footnote1} for $n=1$ one has indeed $p_N= \hbar \alpha^2 w_N$, $w_N$ being
the width near the edge in real space, see \cite{fermions_review}.
\bibitem{fredholm}
We recall that, for a trace-class operator $K(x,y)$ such that ${\rm Tr} K = \int dx K(x,x)$ is well defined, 
$\det(I - K) = \exp{[-\sum_{n=1}^\infty{{\rm Tr \,} K^n}/{n}]}$, where ${\rm Tr}\, K^n = \int dx_1 \cdots \int dx_n K(x_1,x_2) K(x_2,x_3)\cdots K(x_n,x_1)$. The effect of the projector $P_s$ in (\ref{calF}) is simply to restrict the integrals over $s$ to the interval $[s,+\infty )$.
\bibitem{footnotesigma}
The mean of ${\cal F}_4'$ is evaluated as $\approx -1.74$ and
its variance as $\sigma_4^2 \approx 1.18 > \sigma_2^2 \approx 0.81$.
\bibitem{Pearcey}
It can be seen as a higher multicritical generalization of the so-called Pearcey kernel 
\cite{BrezinHikamiSource,TW-Pearcey} which would correspond to a density exponent $1/3$
, see also \cite{MoerbekeGeneralAiry}.
\end{thebibliography}
\end{document}